\documentclass[12pt,preprint]{aastex}
\usepackage{emulateapj5}
\usepackage{apjfonts}

\submitted{To Appear in the Astrophysical Journal
(Received 2002 Apr 5; Accepted 2002 May 8)}

\def\thobs{\theta_{\rm obs}} \def\thjet{\theta_{\rm jet}} 
\def\thmax{\theta_{\rm max}} \def\brel{b_{\rm rel}}

\begin{document}

\title{Orphan Afterglows of Collimated Gamma-Ray Bursts: Rate Predictions and
Prospects for Detection}

\author{Tomonori Totani\altaffilmark{1} and Alin Panaitescu }

\affil{ Princeton University Observatory, Peyton Hall, Princeton, NJ 08544-1001, USA }
\altaffiltext{1}{ %On assignment from 
%On leave from 
Theory Division, National Astronomical Observatory,
Mitaka, Tokyo 181-8588, Japan }

%\date{\today}

\begin{abstract}
We make a quantitative prediction for the detection rate of orphan gamma-ray
burst (GRB) afterglows as a function of flux sensitivity in X-ray, optical,
and radio wavebands, based on a recent model of collimated GRB afterglows. We
find that the orphan afterglow rate strongly depends on the opening angle of
the jet (roughly $\propto \thjet^{-2}$), as expected from simple geometrical
consideration, if the total jet energy is kept constant as suggested by
recent studies. The relative beaming factor $\brel$, i.e., the ratio of all
afterglow rate including orphans to those associated with observable prompt
GRBs, could be as high as $\brel \gtrsim 100$ for searches deeper than $R
\sim 24$, depending on afterglow parameters. To make the most plausible
predictions, we average the model emission for ten sets of afterglow
parameters obtained through fits to ten well-observed, collimated GRB jets,
weighted by the sky coverage of each jet. Our model expectations are
consistent with the results (or constraints) obtained by all past
searches. We estimate the number of orphan afterglows in the first
1500deg$^2$ field of the Sloan Digital Sky Survey (SDSS) to be about 0.2.
The relative beaming factor $\brel$ is rapidly increasing with the search
sensitivity: $\brel \sim 3$ for the SDSS sensitivity to transient objects in
the northern sky ($R \sim 19$), $\sim$14 for the past high-$z$ supernova
searches ($R \sim 23$), and $\sim$50 for the sensitivity of the Subaru
Suprime-Cam ($R \sim 26$). Predictions are made for the current facilities
and future projects in X-ray, optical, and radio bands. Among them, the
southern-sky observation of the SDSS (sensitive to transients down to $R \sim
23$) could detect $\sim$40 orphan afterglows during the five-year
operation. Allen Telescope Array would find about 200 afterglows in a radio
band at $\sim$0.1--1mJy with $b_{\rm rel} \sim 15$.
\end{abstract}

\keywords{gamma rays: bursts --- ISM: jets and outflows --- surveys}

\section{Introduction}

Gamma-ray bursts (GRBs) are now confirmed to be located at cosmological
distances, and they are recognized as the most energetic explosion
in the universe in terms of isotropic equivalent luminosity,
while their origin still remains as a mystery.
The redshift of one GRB 990123 was $z=1.6$, and
its total energy emitted as gamma-rays, estimated by
its energy flux and redshift,  is $E \sim 3 \times 10^{54}$ erg
assuming isotropic radiation (Kulkarni et al. 1999).
This energy is equivalent
to a rest mass energy of $1.7 M_\odot c^2$ and it is almost impossible
to explain by explosions of objects with stellar mass scale. This is why most
researchers in this field now consider that GRBs are likely strongly
collimated, with a typical collimation factor of $4\pi / \Delta \Omega \gtrsim$
100. It is then important to test this hypothesis by observations,
for better understanding of the nature of GRBs.

A direct consequence for such GRB collimation is that the rate of GRB
afterglows may be much higher than those associated with observable prompt
GRBs.  GRBs are believed to be produced by dissipation of kinetic energy of
ultra-relativistic outflow from the central engine with a Lorentz factor of
$\Gamma \sim $100--1000.  The outflow is eventually decelerated by interaction
with interstellar matter, just like supernova remnants,
to produce radiation called afterglows. After $\Gamma$
decreases to $\Gamma \sim \thjet^{-1}$, where $\thjet$ is
the opening angle of the jet, the beaming of radiation is wider than that of
outflow, thus the afterglow becomes observable from directions different from
that of the original gamma-ray radiation.  At the same time, the sideway
expansion increases the opening angle of the jet from 
the original angle of $\thjet$.

Therefore a serendipitous search of GRB afterglows without prompt GRBs is an
important test for the beaming of GRBs (Rhoads 1997).  Several efforts have
been made in various wavebands. Past X-ray surveys set a constraint of $b_{\rm
rel} \lesssim$ several (Grindlay 1999; Greiner et
al. 2000), where $b_{\rm rel}$ is the ``relative'' beaming
factor, i.e., the ratio of all afterglow rate including orphans to the rate of
those with on-axis GRBs.\footnote{ 
We used the term ``relative'' since $b_{\rm rel}$ reflects the
relative beaming of radiation between prompt GRBs and typical afterglows
detectable in an orphan afterglow search.}  Perna \& Loeb (1998) set a crude
upper limit on $b_{\rm rel} \lesssim 1.5 \times 10^3$ by radio source counts.
Schaefer (2002) searched a 264 deg$^2$ field down to $R \sim 21$, but no
afterglow candidate was found. The past survey for high-redshift supernovae
should also give some constraint on the orphan GRB afterglow rate (e.g., Rees
1999).  Recently, Vanden Berk et al. (2002) reported a possible orphan
afterglow found in the initial 1500 deg$^2$ data of the Sloan Digital Sky
Survey (SDSS). However, Gal-Yam et al. (2002) reported that the host galaxy of
the SDSS transient is a highly variable, radio-loud active galactic nuclei
(AGN) and hence it was almost certainly not related to a GRB.

The interpretation of these results is, however, not straightforward.  They
must be compared with a realistic theoretical prediction of $b_{\rm rel}$
which is generally dependent on the waveband and sensitivity of
surveys. Recently, Dalal et al.  (2002) argued, based on a simple analytical
investigation, that the value of $b_{\rm rel}$ is not sufficiently large for
the sensitivity that can be achieved by the current facilities, even if GRBs
are strongly collimated. They also argued that the rate of orphan afterglows
is insensitive to the jet opening angle, and hence they concluded that the
orphan afterglow search hardly constrains the GRB collimation in practice.
What makes the situation even more complicated is the possibility that orphan
afterglows can be produced by other processes rather than the collimation; for
example, a `failed' GRB or a `dirty fireball' may produce afterglows at longer
wavelength while there is no prompt gamma-ray emission (Huang, Dai, \& Lu
2002).

In this paper we present a realistic prediction of orphan afterglow rate as a
function of search sensitivity in various wavebands, based on a popular
afterglow model of collimated GRBs that has been tested against a number of
afterglow observations.  We find that the orphan afterglow rate rapidly
increases with GRB jet collimation roughly as $b_{\rm rel} \propto
\thjet^{-2}$, if GRB jets have roughly constant energies against various
$\thjet$, as suggested by recent studies (Frail et al. 2001; Panaitescu \&
Kumar 2001).  Therefore, the orphan afterglow search could still be powerful
to get information on collimation of GRB jets. Even if the theoretical
prediction is also affected by possible other components of ejecta from
successful or failed GRBs with lower Lorentz factor, it would be useful to
make a prediction for the abundance of orphan afterglows by the `nominal'
(i.e., without any other hypothetical components) jet model of GRB
afterglows, as realistic as possible, since the jet model is a relatively
reliable method of transforming the on-axis light-curves to off-axis ones,
compared with other theoretical possibilities of orphan afterglows. It can be
used as a baseline prediction when one interprets the results of past and
future searches for faint extragalactic transient objects.

To make plausible predictions, we take the observed radio, optical, and X-ray
emission of ten GRB afterglows with a wide variety of estimated jet opening
angle --- 970508, 980519, 990123, 990510, 991208, 991216, 000301c, 000418,
000926, and 010222 --- and calculate their emission as it would be seen at
various angles relative to the jet axis. This is done within the framework of
collimated jets decelerated by the circumburst medium (M\'esz\'aros \& Rees
1997) and undergoing lateral expansion (Rhoads 1999), using the treatment
presented by Panaitescu \& Kumar (2000) and Kumar \& Panaitescu
(2000). With the aid of this model, Panaitescu \& Kumar (2002) have
determined the physical parameters of the ten afterglows by modeling their
broadband emission. We employ these ten sets of
jet parameters to calculate their emission for any off-axis observer
location.  The detection rate as a function of the X-ray/optical/radio search
sensitivity is further obtained by integrating over the viewing angle and
over the GRB rate history in the universe.

The paper will be organized as follows. \S \ref{section:model}
reviews the model of collimated afterglows used here, and some discussions
are presented on the physics of collimated afterglows concerning the 
detectability of orphan afterglows. \S \ref{section:formulation}
is for the formulation of orphan afterglow rate calculation.
The results are presented in \S \ref{section:results}; they are
compared with various past search results and predictions for the current
and future facilities are made. Discussions are given in \S
\ref{section:discussion}, including comparison with previous studies,
another possible picture of GRB jets, and caveats of our predictions.
Our main conclusions are presented in \S \ref{section:conclusions}.

\section{The Afterglow Model of Collimated GRBs}
\label{section:model}

 The dynamics of a GRB jet is calculated by tracking its total energy
(allowing for radiative losses) and mass. The jet spreads laterally at the
sound speed, and it is assumed to be uniform within its aperture
(i.e. the same energy per solid angle in any direction) and to have sharp
boundaries.  The afterglow synchrotron emission and the radiative losses are
calculated from the strength of the magnetic field in the shocked gas and the
power-law distribution of shock-accelerated electrons, taking into account
their radiative (synchrotron and inverse Compton) cooling.  To obtain observer
light-curves, the jet emission is integrated over its surface, allowing for
the differential relativistic beaming and the spread in the photon arrival
time across the jet surface.

 The afterglow jet model has three parameters that determine the jet dynamics 
(the initial jet energy $E_{\rm jet}$, initial jet half-angle $\thjet$,
and external particle density $n_{\rm ext}$), and three parameters pertaining
to the microphysics of shocks (the fraction of the post-shock energy density
in magnetic fields $\epsilon_B$, the fractional energy of the least energetic
injected electrons $\epsilon_e$, and the power-law index $p$ of the shock-accelerated
electron energy distribution). For hard electron distributions with $p < 2$,
two other parameters become relevant: the total fraction of the internal energy 
of the shocked fluid imparted to electrons, which sets a ``cut-off" of the electron
distribution at high electron energies, and the larger (than 2) power-law index 
of the electron index above the cut-off. The passage of the spectral break
associated with the electron distribution cut-off through the optical domain
may yield a light-curve break, as suggested by the sharp break seen in 000301c
and the steep decay of the optical emission of 991208.

\begin{figure}
\plotone{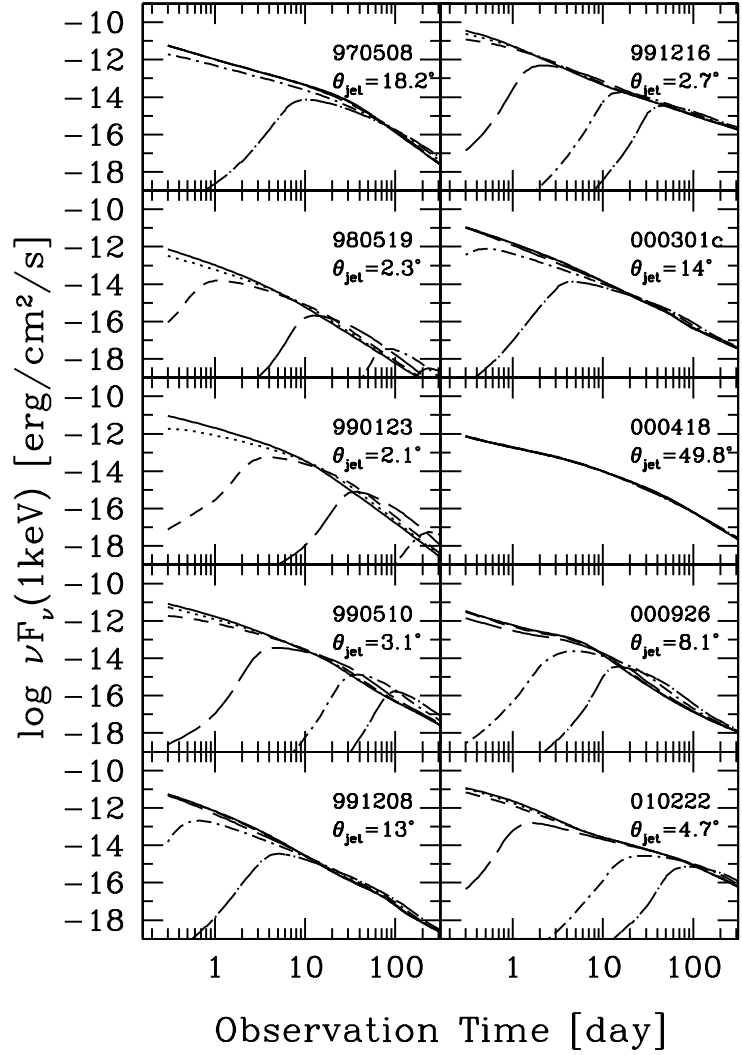}
%\plotone{f1.eps}
\caption{Light curves of off-axis GRB afterglows in X-ray band (1keV)
for the ten well-observed GRBs (name and the opening angle
of the jet indicated in each panel). Note that the distance is
assumed to be $z=1$ for all GRBs, for comparison.
The light curves are shown for different viewing angles from
the center of the jet, as $\thobs$ = 1, 3, 5, 10, 20,
and $30^\circ$ for the solid, dotted, short-dashed, long-dashed,
short-dot-dashed, and long-dot-dashed lines, respectively.
}
\label{fig:lc-X}
\end{figure}

\begin{figure}
\plotone{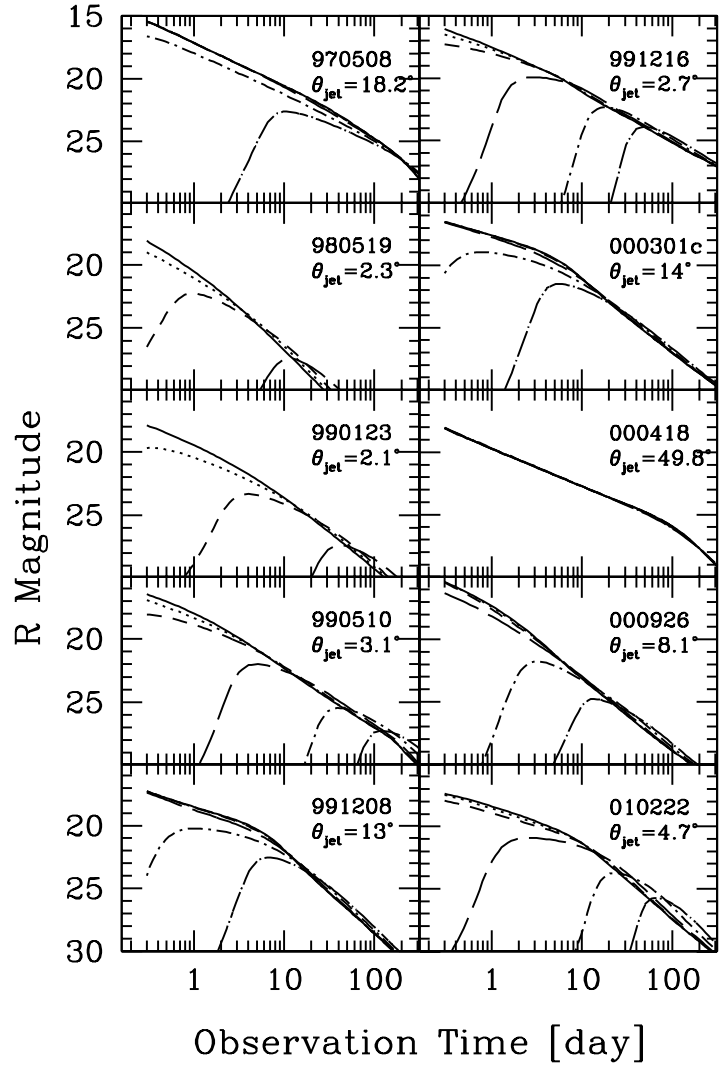}
%\plotone{f2.eps}
\caption{The same as Fig. \ref{fig:lc-X}, but for the optical ($R$)
band.
}
\label{fig:lc-R}
\end{figure}

\begin{figure}
\plotone{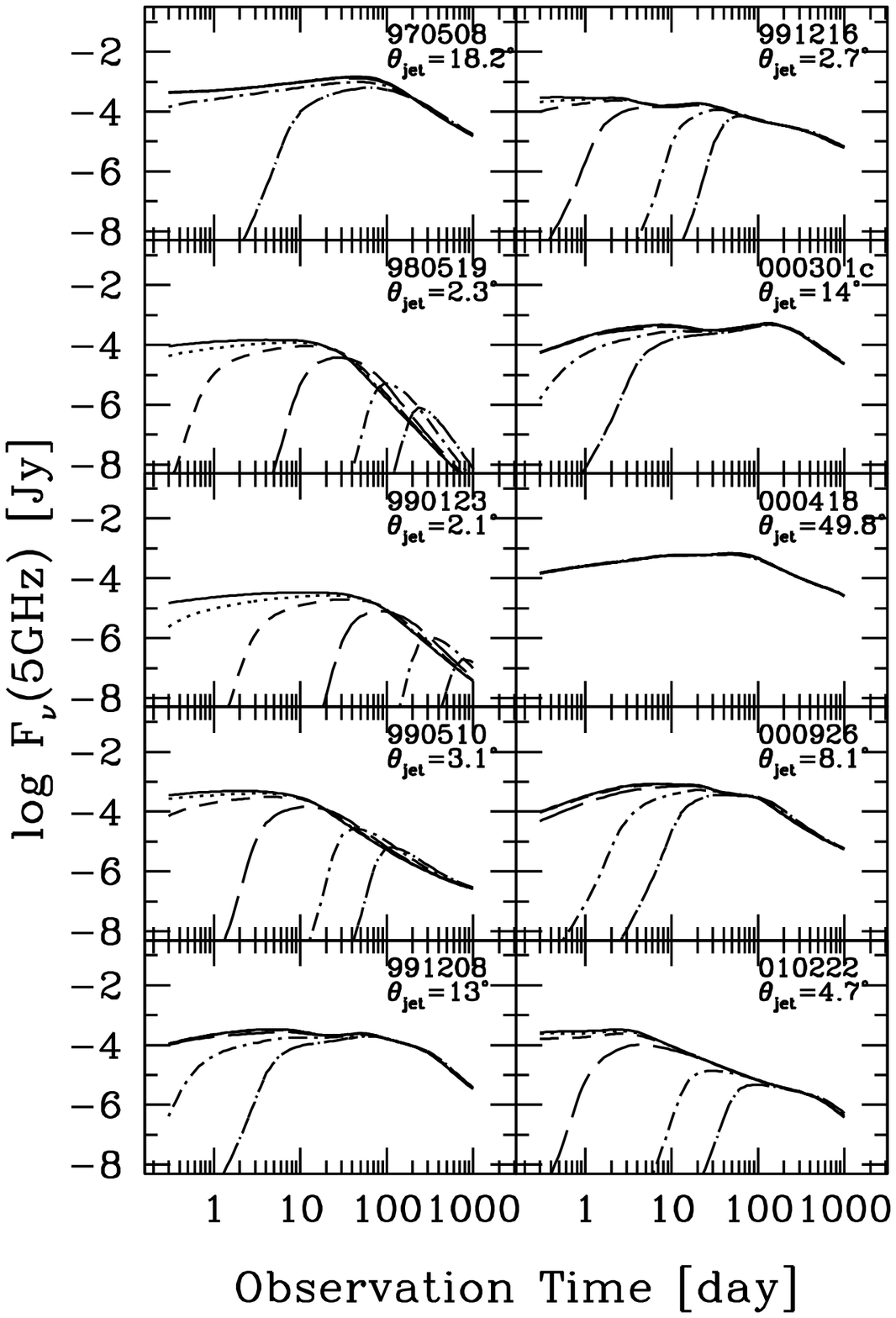}
%\plotone{f3.eps}
\caption{The same as Fig. \ref{fig:lc-X}, but for the radio (5GHz)
band.
}
\label{fig:lc-radio}
\end{figure}

 Figures \ref{fig:lc-X}--\ref{fig:lc-radio} show the X-ray, optical, and radio
afterglow light curves for the above ten jets, as they would be seen from
various viewing angles relative to the center of the jet, $\thobs$. It can be 
seen that the behavior of off-axis light curves is very
different for different sets of model parameters, even when the opening angle
of the jet, $\thjet$, is similar. As illustrated in Figure
\ref{fig:lc-R}, `the best case', i.e., the highest rate for orphan afterglow
searches is expected for GRB 991216 ($\thjet = 2.7^\circ$), whose
peak of optical light curves is at $R \sim 24$ (when it is placed at $z=1$)
for an observer
located with an angle $\thobs = 30^\circ$. If one searches orphan
afterglows with a sensitivity better than this, one could find them at more
than 100 times higher rate than that expected from spherical GRBs (i.e.,
$b_{\rm rel} > 100$). The other extreme is the cases of GRB 980519 or
990123. Even though their jet opening angle $\thjet$ is not much
different from that of GRB 991216, the peaks of optical light curves are
difficult to detect when an observer is located at $\thobs \gtrsim
10^\circ$. In these cases the relative beaming factor is at most $b_{\rm rel}
\lesssim 10$.

 The dependence of $\brel$ on the jet initial opening (and other jet
properties), as well as the reason for which $b_{\rm rel}$ is maximal for the
991216 afterglow, can be obtained using a simplified model where all the
afterglow emission originates from a single point on the jet axis (Dalal et
al. 2002, Granot et al. 2002) whose dynamics is that of an expanding jet
(Rhoads 1999). For narrow ($\thjet \ll 1$ rad), relativistic jets and a
homogeneous medium, it can be shown that the jet Lorentz factor $\gamma$ is
\begin{equation}
\label{gamma}
 \gamma(t_0) = \thjet^{-1} \left( \frac{t_0}{t_{j,0}} \right)^{-n}
               \quad  \left\{
 \begin{array}{ll}
    n = 3/8 \;,\; t_0 < t_{j,0} \\
    n = 1/2 \;,\; t_0 > t_{j,0}  
 \end{array} \right. \;,
\end{equation}
where the subscript "0" indicates the time (since the prompt gamma-ray
emission) measured by an on-axis observer ($\thobs = 0$), $t_{j,0}$
being the "jet-break time", i.e$.$ the time when $\gamma = \thjet^{-1}$ 
and the on-axis observer "sees" the entire jet surface and
a break in the light curve.  Taking
into account that the photon arrival time for an arbitrary observer location
$\thobs$ is given by $dt/dt_0 = 2\gamma^2 (1 - \beta \cos \thobs)$, 
$\beta$ being the jet speed in units of the speed of light, 
equation (\ref{gamma}) gives the jet Lorentz factor as a function
of the time $t$ when photons arrive at $\thobs$:
\begin{equation}
\label{Gamma}
 \frac{ \gamma(t) }{ f^n(\gamma) } = \frac{ \thjet^{-1} }{ f^n (\thjet^{-1}) } 
               \, \left( \frac{t}{t_j (\thobs)} \right)^{-n} \;
\end{equation}
where 
\begin{equation}
 f(\gamma) \equiv \cos\thobs + 8[2-\ln(\min\{1,\gamma\thjet\})] 
            \,\gamma^2 \sin^2 (\thobs/2) \;.
\end{equation}
The above $f(\gamma)$ relates the photon arrival time at $\thobs$ to that 
for an on-axis observer: $t = f(\gamma) t_0$, therefore, in equation 
(\ref{Gamma}), $t_j (\thobs)=f(\thjet^{-1}) t_{j,0}$.
For an observer located well outside the jet opening ($\thobs \gg \thjet$) 
the received jet emission is Doppler boosted in frequency by the same factor 
${\cal D} = [\gamma (1 - \beta \cos \thobs)]^{-1}$, its intensity being
relativistically enhanced by ${\cal D}^3$. Thus the received afterglow flux 
$F (\nu,t)$ at $\thobs$ is related to that seen by an on-axis observer, 
$F_0 (\nu,t_0)$, through
\begin{equation}
\label{Fnu}
 F (\nu,t) = g^{-3}(\gamma) F_0 (g\nu, t/f) \;, 
\end{equation}
where
\begin{equation}
 g(\gamma) \equiv \cos \thobs + 4 \gamma^2 \sin^2 (\thobs/2) \;.
\end{equation}

 The observed GRB afterglows have power-law optical spectra and light-curves:
$F_0 (\nu, t_0) \propto \nu^{-\delta} t_0^{-\alpha}$ (for the above-mentioned
ten afterglows, $\delta \in [0.6,1.5]$, $\alpha \equiv \alpha_1 \in
[0.7,1.7]$ at $t_0 < t_{j,0}$, and $\alpha \equiv \alpha_2 \in [1.6,3.0]$ at
$t_0 > t_{j,0}$). Therefore equation (\ref{Fnu}) leads to
\begin{equation}
\label{fnu}
 F (\nu, t) \propto f^\alpha (\gamma) [g(\gamma)]^{-(3+\delta)} t^{-\alpha} \;,
\end{equation}
with $\gamma(t)$ given by equation (\ref{Gamma}).

 For $t < t_j$, when $\gamma \thjet > 1$, the above equations yield
$\gamma \propto t^{-3/2}$ and $F (\nu,t) \propto \gamma^{2(\alpha_1-\delta-3)}
t^{-\alpha_1} \propto t^{3(3+\delta)- 4\alpha_1}$ in the $\thobs \ll 1$ limit, 
i.e$.$ a light-curve with a sharp rise.  At $t > t_j$ and before the
time $t_p$ when $\gamma(t_p) = \thobs^{-1}$, $\gamma \propto \exp\{-2t/t_j\}$ 
and $F (\nu,t) \propto \exp\{4(3+\delta-\alpha_2)(t/t_j)\}$, thus the light-curve 
continues to rise. At $t > t_p$ the $f$ and $g$ functions asymptotically approach
unity, so that the evolution of the jet Lorentz factor and the afterglow 
light-curves become those for an on-axis observer: $\gamma \propto t^{-1/2}$ and 
$F (\nu, t) \propto t^{-\alpha_2}$, respectively.
Therefore the afterglow light-curve for $\thobs$ peaks around the time $t_p$ when 
$\gamma(t_p) \thobs = 1$.
Taking into account that $t_p = f[\gamma(t_p)] t_{p,0}$, where $t_{p,0} \simeq 
(\thobs/\thjet)^2 t_{j,0}$ (from eq$.$ [\ref{gamma}])
is the corresponding photon arrival time for an on-axis observer and
$f(\thobs^{-1}) \simeq 5 + 2\ln(\thobs/\thjet)$ in the 
$\thobs \ll 1$ limit, it follows that the afterglow light-curve seen by 
an observer at $\thobs$ peaks at
\begin{equation}
\label{tpeak}
 t_p = \left( 5 + 2\ln \frac{\thobs}{\thjet} \right)
        \left(\frac{\thobs}{\thjet} \right)^2 t_{j,0} \;.
\end{equation}
In the same $\thobs \ll 1$ limit, $g[\gamma(t_p)] \simeq 2$, 
therefore equations (\ref{Fnu}) and (\ref{tpeak}) give for the peak flux 
\begin{equation}
\label{fpeak}
 F (\nu, t_p) = 2^{-(3+\delta)} \left( \frac{\thobs}{\thjet} \right)
                 ^{-2\alpha_2} F_0 (\nu, t_{j,0}) \;.
\end{equation}
It may not be easy to infer $t_p$ for an observed orphan afterglows,
since we do not know the precise time of the prompt burst time.
However, if there are enough number of data points to construct
a light curve, it may be possible to infer $t_p$ only from orphan
afterglows. Then,
together with an empirical relation between the jet-break time $t_{j,0}$ and
flux $F_0 (\nu, t_{j,0})$, calibrated with the afterglows which were seen
on-axis (i.e$.$ preceded by a GRB), equations (\ref{tpeak}) and (\ref{fpeak})
can be used to determine the relative observer location $\thobs/\thjet$,
$t_{j,0}$, and $F_0 (\nu, t_{j,0})$ from the inferred peak time $t_p$ and 
observed flux $F (\nu, t_p)$ of an orphan GRB afterglow.

 From equation (\ref{fpeak}) it follows that only jets seen at an angle less
 than
\begin{equation}
 \thmax = \left[ 2^{-(3+\delta)} \frac{ F_0 (\nu, t_{j,0}) }{ F_{\rm lim} } 
          \right]^{1/(2\alpha_2)} \, \thjet \; ,
\label{eq:theta-max}
\end{equation}
can be detected above a given detection threshold $F_{\rm lim}$. Therefore the
relative beaming factor is $b_{\rm rel} = (\thmax/\thjet)^2 \propto
[2^{-\delta} F_0(\nu,t_{j,0})]^{1/\alpha_2}$. The $b_{\rm rel}$ that can be
inferred for each afterglow shown in Figure \ref{fig:lc-R}, for a given
threshold, is consistent with this estimation of $b_{\rm rel}$.  Among the
set of ten afterglows, 991216 has one of the largest jet-break fluxes $F_0
(\nu, t_{j,0})$ (note that for this afterglow $t_{j,0} \lesssim 1$ day, the
sharper break seen in Figure \ref{fig:lc-R} at several days being due to the
passage through the optical band of a spectral break), hardest optical
spectra, and shallowest decays after the jet-break time. For these reasons,
its $b_{\rm rel}$ is the largest in our sample of afterglows, and GRB991216
dominates the detection probability of orphan afterglows.

\section{Formulations for the Rate Calculation}
\label{section:formulation}

First we calculate the expected number of orphan afterglows per unit solid
angle $N_{\rm exp}(F_{\rm lim})$, which can be detected by a snapshot
observation with a given sensitivity $F_{\rm lim}$.  This quantity directly
provides the detectability of afterglows in a search where transient objects
are found by comparison of several snapshot images taken at time intervals
longer than the duration $T$ over which the afterglow is brighter than the
sensitivity.  On the other hand, the detection rate in a survey of consecutive
monitoring with a time scale longer than $T$ can be estimated by $R_{\rm exp}
\sim N_{\rm exp} /T$ per unit solid angle and observation time.

The afterglow flux at any frequency and
observer's time viewed by an observer with arbitrary $\thobs$
is calculated with the model presented in the previous section.
From there we can calculate
the time duration for an observer, $T(x, z, \thobs, F_{\rm lim})$,
during which an afterglow is brighter than a given sensitivity, 
where $z$ is redshift and $x$ symbolically
represents a specific type of GRB afterglows.
Then, the expected number $N_{\rm exp}$ per all sky can be written as:
\begin{eqnarray}
N_{\rm exp} (F_{\rm lim})
= \int dx \int dz \int_0^\pi
d\thobs
\frac{\sin \thobs}{2}  \nonumber \\
\times \frac{dV}{dz} \frac{R_{\rm GRB}(x, z)}{1+z} 
 T(x, z, \thobs, F_{\rm lim}) \ ,
\label{eq:Nexp1}
\end{eqnarray}
where $(dV/dz)$ is the standard comoving volume element in all sky,
and $R_{\rm GRB}(x, z)$ is the comoving GRB rate density of type $x$.
It should be noted that $R_{\rm GRB}$ is the true rate of GRBs
including unobservable ones whose jet are not directed to us.

As shown in the previous section, the detectability of orphan afterglows
sensitively depends on afterglow parameters. In order to take this into
account and make the most plausible prediction in an empirical way, we
replace the integration over $x$ by the sum of the ten well-observed
afterglows presented in the previous section.  We assume that all the
population of observable GRBs is represented by those ten bursts with an
equal weight, and it is independent of redshift. If detection of GRBs is flux
limited, then a simple sum of the ten observed bursts having very different
gamma-ray luminosities might induce some bias from the true
population. However, GRBs are generally bright enough to be detected even at
very large cosmological distances, and hence the effect of flux limit is
expected to be insignificant.  In addition, the redshifts of the ten GRBs are
in a rather small range.  Therefore, we expect that the above treatment is
not unreasonable.  On the other hand, it should be noted that we must take
into account the estimated opening angle of jets; the GRB rate inferred from
the observed number of GRBs assuming isotropic emission ($R_{\rm iso}$) is
not the true GRB rate, but it should be multiplied by the inverse of sky
coverage of gamma-ray emission, $2/(1-\cos \thjet)$, to obtain $R_{\rm
GRB}$. \footnote{ Here, we have assumed that the opening angle of gamma-ray
emission is the same with that of the jet inferred from the afterglow
fitting, but this is not necessarily true. We will discuss this point later
in \S\ref{section:discussion-caveats}.}  Then, the integration over $x$ in
equation (\ref{eq:Nexp1}) should be replaced by the sum of $N_{\rm GRB}$ (=10)
GRBs, as:
\begin{eqnarray}
N_{\rm exp} (F_{\rm lim})
= \sum_{i = 1}^{N_{\rm GRB}} \frac{1}{N_{\rm GRB}}
\int dz \int_0^\pi
d\thobs
\frac{\sin \thobs}{2}  \nonumber \\
\times
\frac{dV}{dz} \frac{R_{\rm iso}(z)}{1+z}  
\left( \frac{1 - \cos \theta_{\rm jet, \it i}}{2} \right)^{-1}
T(i, z, \thobs, F_{\rm lim}) \ .
\label{eq:Nexp2}
\end{eqnarray}
It should be noted that the ten GRBs chosen by Panaitescu \& Kumar (2002)
are well observed, with good temporal coverage at various frequencies, and
showing a light curve break. This suggests that there may be a systematic
bias toward strongly collimated GRBs in our selection. 
We note that the ten GRB
afterglows used here represent about a third of all bursts for which optical
afterglows were found. This indicates that the formulation above
overestimates by a factor of up to three the real orphan afterglow rate if
the other 2/3 GRB afterglows are isotropic and hardly contribute to $N_{\rm
exp}$.  This is probably not much larger than other model uncertainties.
Therefore we believe that the ten GRBs represent the whole GRB population
reasonably well.

We assume that $R_{\rm iso}(z)$ traces the cosmic star formation
history (Totani 1997; Wijers et al. 1998). Based on the recent
observational estimates (see, e.g., Totani \& Takeuchi (2002) for a  summary of
recent studies on this issue), we assume that $R_{\rm iso}(z)
\propto (1+z)^{3.77}$ at $0 < z < 1$, and $R_{\rm iso}(z)$ is constant
at $z>1$, when the Einstein-de Sitter universe is adopted. We set an upper
cut-off for redshift as $z_u = 5$.
The normalization of this rate density is fixed at $z=0$, as
$R_{\rm iso}(0) = 2\times 10^{-10} (h/0.7)^3 \ \rm yr^{-1} Mpc^{-3}$
where $h = H_0/(\rm 100km \ s^{-1}Mpc^{-1})$, based on the fitting to
the observed flux distribution (i.e., log$N$-log$P$ relation)
of the BATSE GRBs (Totani 1999; Schmidt 1999). Throughout this paper
we adopt a standard $\Lambda$-dominated universe with
($h, \Omega_0, \Omega_\Lambda)$ = (0.7, 0.3, 0.7) and the
cosmology-dependence of $R_{\rm iso}(z)$ is corrected appropriately.

\section{Results}
\label{section:results}

\begin{figure}
\plotone{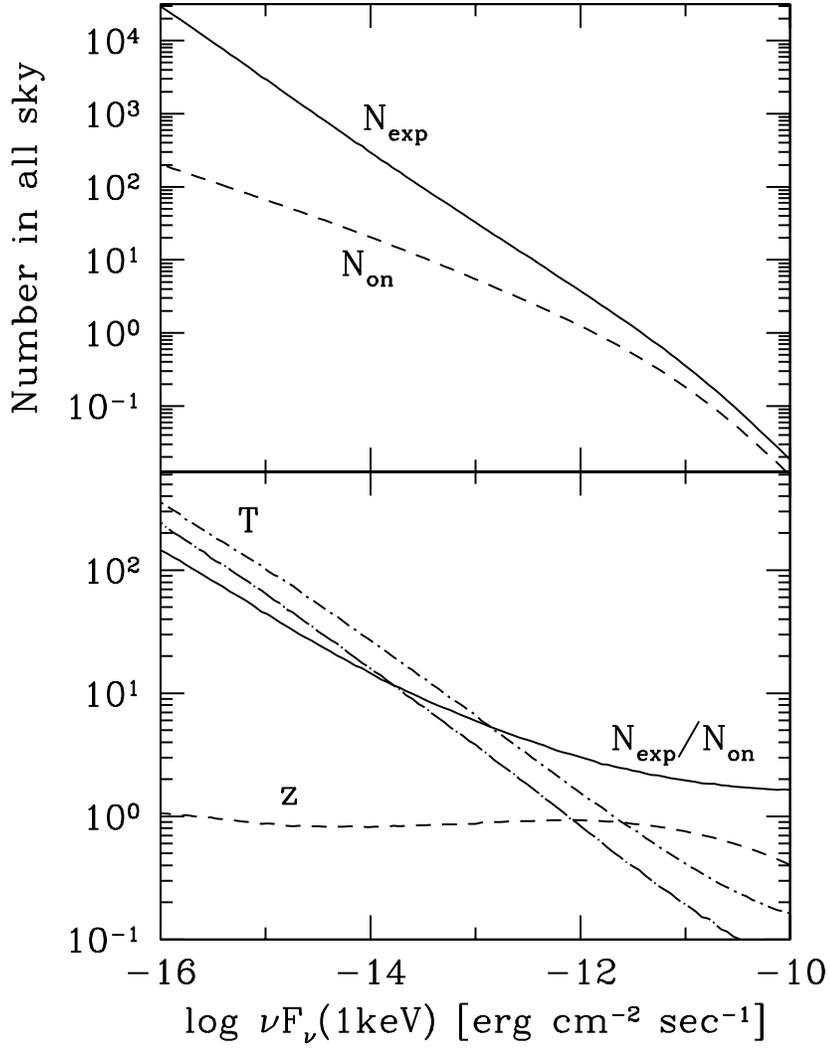}
%\plotone{f4.eps}
\caption{Expected number of orphan afterglows detectable by a snapshot 
observation per all sky, in the
X-ray band (1keV) as a function of sensitivity limit (upper panel). 
The solid line ($N_{\rm exp}$)
is for all afterglows including orphans,
while the dashed line ($N_{\rm on}$)
is for on-axis afterglows which are associated
with observable prompt GRBs, i.e., those with $\thobs < \thjet$. 
In the lower panel, the ratio of $N_{\rm exp} / N_{\rm on}$, 
the mean values of redshift ($10^{\langle \log z \rangle}$, 
dashed line)
and time duration ($T$ in days) over which the flux is
above a given sensitivity are shown. For the time duration,
two different means of $10^{\langle \log T \rangle}$ (short-dot-dashed) and
$\langle 1/T \rangle^{-1}$ (long-dot-dashed) are shown.
}
\label{fig:rate-X}
\end{figure}

\begin{figure}
\plotone{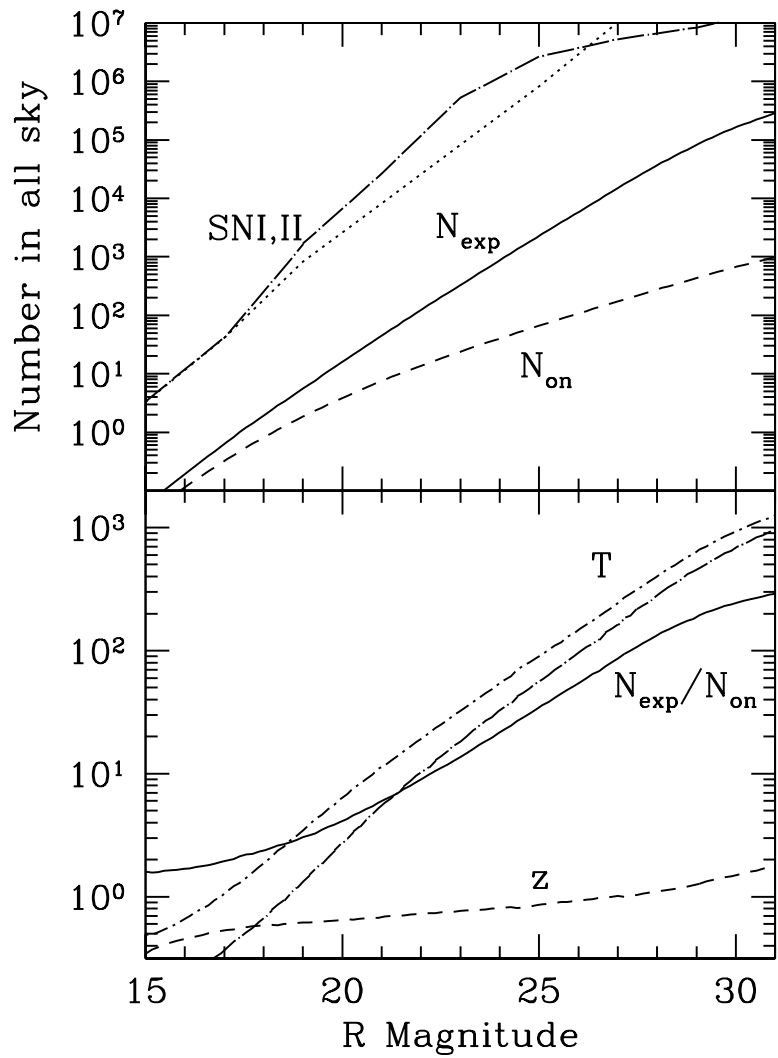}
%\plotone{f5.eps}
\caption{The same as Fig. \ref{fig:rate-X}, but for the optical ($R$)
band. In addition, the expected number of supernovae of type Ia (dot-dashed
line) and type II (dotted line) calculated by Woods \& Loeb (1998) 
are also shown in the upper panel.
}
\label{fig:rate-R}
\end{figure}

\begin{figure}
\plotone{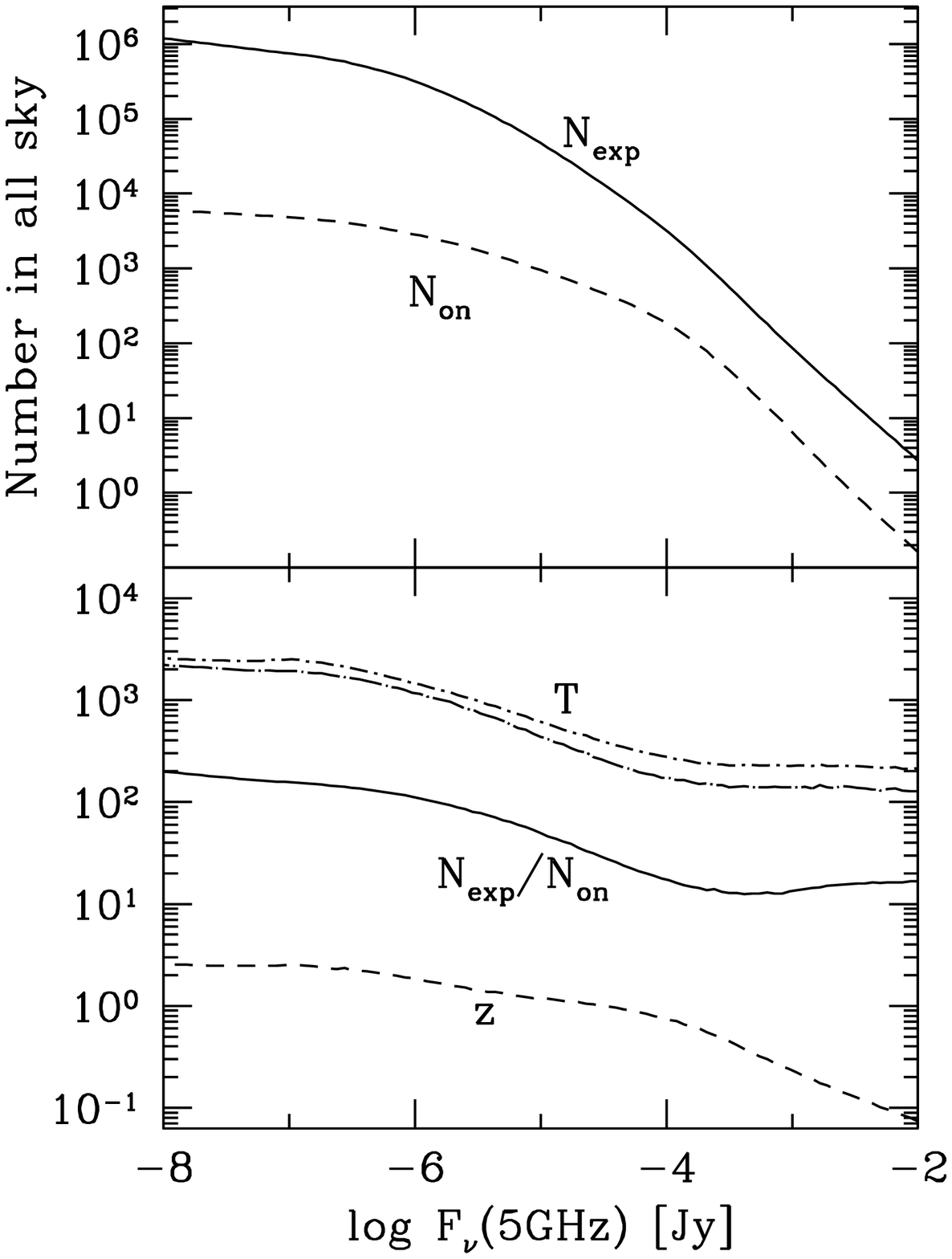}
%\plotone{f6.eps}
\caption{The same as Fig. \ref{fig:rate-X}, but for the radio (5GHz)
band.
}
\label{fig:rate-radio}
\end{figure}

Figs. \ref{fig:rate-X}--\ref{fig:rate-radio} show the result of calculation
for $N_{\rm exp}$, as a function of sensitivity in X (1 keV), optical ($R$),
and radio (5 GHz) wavebands, respectively. We also show the rate expected
only from on-axis afterglows, i.e., associated with prompt GRBs. This rate,
$N_{\rm on}$, is obtained by replacing the range of integration over $\thobs
= 0-\pi$ by $\thobs = 0-\thjet$ in eq (\ref{eq:Nexp2}).  The relative beaming
factor, $b_{\rm rel} \equiv N_{\rm exp}/ N_{\rm on}$ is shown in the lower
panel. The mean values of redshift and $T$ are also shown; here we calculated
average of $\log z$ and $\log T$ because of considerable scatter of these
values at a fixed sensitivity. We also give another mean time,
\begin{equation}
\left\langle \frac{1}{T} \right\rangle^{-1} \equiv 
\left( \frac{\int dN_{\rm exp} / T}{ N_{\rm exp}} \right)^{-1}\ ,
\end{equation}
where integration with $dN_{\rm exp}$ represents that of
eq. (\ref{eq:Nexp2}). We give this quantity because we can translate $N_{\rm
exp}$ into detection rate per time in a consecutive monitoring observation by
$R_{\rm exp} = N_{\rm exp} \langle 1/T \rangle$, taking into account the
dispersion of $T$. For comparison, $N_{\rm exp}$ for type Ia and II
supernovae calculated by Woods \& Loeb (1998) for the optical band 
is also shown in Fig. \ref{fig:rate-R}.

A clear trend can be seen in all wavebands; the relative beaming factor
$b_{\rm rel}$ is rapidly increasing with the sensitivity limit, while the
mean redshift is not.  These results reflect the fact that intrinsically
faint afterglows at small distances with large $\thobs$ become dominant when
the search sensitivity is improved.  For orphan afterglow search with $b_{\rm
rel} \gtrsim 10$, the required sensitivities are $\nu F_\nu \sim
10^{-14}$--$10^{-13} \ \rm erg \ cm^{-2} s^{-1}$ and $R \sim 22$, in X-ray
(1keV) and optical bands, respectively.  In the radio bands, $b_{\rm rel}$
exceeds 10 in all flux range shown here, while a sensitivity of $\sim
10\mu$Jy is necessary to increase $b_{\rm rel}$ to more than 100.

\begin{figure}
\plotone{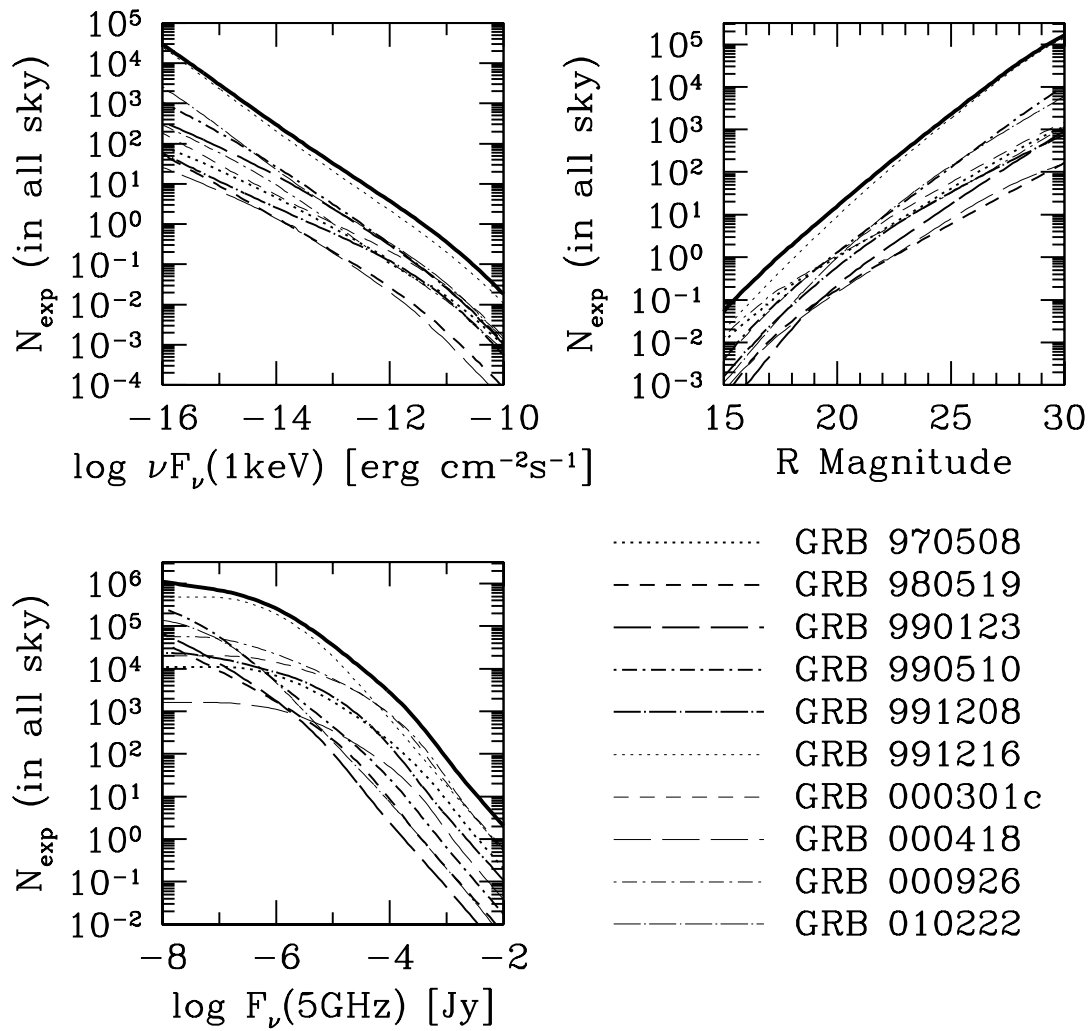}
%\plotone{f7.eps}
\caption{Contribution of each of the ten GRB afterglows to the
detection rate $N_{\rm exp}$ (shown by the thick solid line),
in the three wavebands. The line markings for the ten afterglows
are shown in the figure.
}
\label{fig:rate-each}
\end{figure}

Figure \ref{fig:rate-each} shows the contribution from each of the
ten GRBs to the total rate of orphan afterglows. As expected,
GRB 991216 dominates in most cases, while others are important as well
in bright flux range in radio bands. It is interesting to check
how the prediction is changed when the dominant GRB 991216 is removed
in the sample for averaging. In the optical band, the mean rate
is reduced by a factor of 2.2 and 4.5 at the search sensitivities of
$R$ = 20 and 25, respectively, when GRB 991216 is removed.

\section{Comparison with Past Searches, and Prospects for Future Projects}

In this section we calculate the number of orphan afterglows expected in
several past surveys and compare them with the reported results. We also make
predictions for the number of orphan afterglows expected in future surveys.
The summary of our results is given in Table \ref{table:projects}.

\subsection{X-ray Observations}

There are two papers which constrained the orphan afterglow rate in X-ray
band: Grindlay (1999) by the {\it Ariel} 5 survey with a sensitivity of $\sim
10^{-10} \ \rm ergs \ cm^{-2} s^{-1}$ (2--10keV), and Greiner et al. (2000) by
the ROSAT All Sky Survey (RASS) which is sensitive to $\sim 10^{-12} \ \rm
ergs \ cm^{-2} s^{-1}$ (0.1--2.4keV). Figure \ref{fig:rate-X} shows that the
relative beaming factor $b_{\rm rel}$ is about 1.6 and 3.0 for the
sensitivities of the {\it Ariel} 5 and RASS, respectively. Considering this
small number, the negative result of Grindlay (1999) seems consistent with our
expectation. The exposure of RASS is $76,435 \rm \ deg^2 \ days$, and using
$\langle 1/T \rangle^{-1} \sim $ 0.8 day at the RASS sensitivity, this is
equivalent to a snapshot observation of $\sim 94,000 \rm \ deg^2$. Our model
predicts about 8 GRB afterglows for this sky coverage, among which $\sim$3
are expected to be on-axis.  Greiner et al. found 23 candidates of afterglows,
but they argued that the bulk of these are likely to be nearby flaring stars.
If one removes those which are suspected as flaring stars, the remaining
candidates are at most $\sim$10. It is interesting that this number is
very close to our expectation.

The advanced satellites such as Chandra or XMM-Newton have a typical
sensitivity limit of $\sim 10^{-15} \ \rm erg \ cm^{-2} sec^{-1}$ for point
sources, with a field of view of $\sim 10^3 \ \rm arcmin^2$. At such a
sensitivity, the relative beaming factor is increased to $b_{\rm rel} \sim$
40.  Our model predicts that the probability of finding an orphan afterglow
in the field of view of one snapshot observation is $\sim$0.02. This is
apparently small, but accumulation of archive data might be useful to search
and constrain orphan afterglows in the future, though discrimination from
flaring stars could be a major problem again.

\subsection{Optical Observations}
\subsubsection{Past Optical Surveys}

Schaefer (2002) searched orphan afterglows in a 264 square degree field with
a sensitivity of $R=21$. The field was examined nightly, with a total
duration of 33 days.  He found no afterglow candidates.  By using $\langle
1/T \rangle^{-1} \sim 5.1$ days of our prediction, this observation
corresponds to a snapshot observation of $264 \times 33 / 5.1 \sim 1,500$
square degree.  Our model predicts that about two orphan afterglows are
expected, which is $b_{\rm rel} \sim 6.2$ times higher than the case of no
beaming. Therefore, our model is marginally consistent with his result. On
the other hand, there is a good chance to detect orphan afterglows by
continued search and the effort in this direction is encouraged.

High redshift supernovae have been intensively searched down to $R \sim 23$,
for the purpose of determination of cosmological parameters
(e.g., Schmidt et al. 1998; Perlmutter et al. 1999).
The total exposure is about
a few tens of ``square degree years'' (Rees 1999). They are normally
searching supernovae with two images separated by one month,
which is longer than the characteristic afterglow time scale
at this sensitivity ($\langle 1/T \rangle^{-1} \sim 18$days). Therefore,
the exposure is equivalent to 12 times searches in a field of a few tens
of square degree. Then we estimate the expected detection number is
about 1.6 in the past, which is  $b_{\rm rel} \sim 14$
times higher than the case of no beaming. It should be noted that
the past supernova searches (including that of Schaefer)
are not perfectly suitable to orphan
afterglow search. Supernovae are found only in one bandpass (i.e., no
color information), and spectroscopic follow-up is done only for
a small part of them showing good properties for the cosmological use.
It should also be noted that they sometimes find ``mystery objects'',
which decay faster than supernovae and have no host
galaxies (Schmidt et al. 1998). These objects might be orphan afterglows.

\subsubsection{Sloan Digital Sky Survey}

Recently Vanden Berk et al. (2002) reported an interesting, highly luminous
transient object which might be an orphan afterglow, found in the initial 1500
deg$^2$ data of the Sloan Digital Sky Survey. However, Gal-Yam et al. (2002)
claimed that the host galaxy of the SDSS transient is in fact an unusual
radio-loud AGN showing strong variability. Therefore, it seems very likely
that the SDSS transient was an AGN flare, though the possibility of an orphan
afterglow cannot be completely rejected.  Here we give an estimate of expected
number of orphan afterglows in the SDSS data taken so far.  Vanden Berk et
al. (2002) searched transient objects by the flux difference between imaging
observations and later spectroscopic observations. Although the SDSS imaging
observation has a sensitivity of $r' \sim 23$, spectroscopy is not performed
for all objects; the object was selected as a quasar candidate by its
colors. For low redshift quasars whose colors are similar to GRB afterglows,
the sensitivity is $i' \sim 19$ (Stoughton et al. 2002).
Since $R - i' \sim 0$ for typical GRB afterglows, we can
consider that the sensitivity of the search made by Vanden Berk et al. is $R
\sim 19$, with a survey area of 1500 deg$^2$. Then, we found that the expected
number of orphan afterglows to be $\sim 0.2$, which is $b_{\rm rel} \sim 3.0$
times higher than the case of no beaming.  This number is small, but not
extremely small compared with one detection.  (Note, however, that the number
becomes smaller by a factor of 10 when we apply the brightness of the SDSS
transient, $R \sim 17$, as the search sensitivity.)  Our model predicts the
mean redshift at this sensitivity to be $10^{\langle \log z \rangle} \sim
0.6$, and the 1$\sigma$ dispersion around this mean is $\sigma_{\log
z}\sim$0.5. Therefore the redshift $z=0.385$ of the host galaxy well falls in
the plausible range. The characteristic time scale predicted by the model,
$10^{\langle \log T \rangle} \sim $ 3.5 days with $\sigma_{\log T} \sim 0.7$
is also consistent with the modest variability shown by the SDSS objects
during the first two observations separated by 2 days. Therefore, it is not
unreasonable even if the SDSS transient is an orphan afterglow, although the
strong variability of the host galaxy indicates that the transient is much more
likely an AGN flare.  More statistics is obviously needed for stronger
conclusions.

When the SDSS project is completed, the covering area in the northern sky will
be increased to 10,000 deg$^2$.
Then we expect about 1.3 orphan afterglows, which seems still
small to test our model with sufficient statistics by the future data.  On the
other hand, the SDSS has another observing mode in the southern sky, observing
a 225 deg$^2$ field repeatedly to achieve a much deeper sensitivity limit than
the northern sky (e.g., Ivezi\'c et al. 2000).  These southern data can be
used for transient object search with a sensitivity of $R \sim 23$, which is
significantly deeper than the search using spectroscopic data. The southern
sky field is typically observed about four times in a year with a separation
longer than one month.  Then, at the end of the project after the planned five
year operation, we expect an effective survey area of $225 \times 4 \times 5
\sim 4500$ deg$^2$ (\v{Z}. Ivezi\'c 2002, a private communication). Then we
expect about 36 orphan afterglows, which is $b_{\rm rel} \sim 14$ times higher
than the case of no beaming.  This number suggests that we may detect a
statistically meaningful number of orphan afterglows. Although spectroscopic
information will not be immediately available, follow-up observations for host
galaxies will give redshift information. The five-bands photometry data will
also help to discriminate the orphan candidates from other transient objects
(Rhoads 2001). Contamination of highly variable AGNs, as in the case of
the SDSS transient of Vanden Berk et al. (2002), can be removed by close
examination of host galaxies in X-ray and/or radio bands and past
records. (It should be noted that the host galaxy of the SDSS transient
reported by Vanden Berk et al. has a strong radio emission that cannot
be explained by star formation activity.)
Finally, a significant part ($\sim$ 10--40\%) of the northern sky
will also be observed more than two times, because of overlaps of
field-of-views. The time intervals of these repetitions are not simple but
depending on the survey schedule. These data can also potentially be used for
orphan afterglow searches.

\subsubsection{OGLE III}
The Optical Gravitational Lensing Experiment in the third phase
(OGLE III, Udalski et al. 2002) has started its operation, and
during the LMC and SMC season, it will cover a 85 square degree field 
every night or every second night for half a year. The limiting
magnitude is about $I=19.5$ at S/N=10 for a typical exposure of 2 minutes,
corresponding $R \sim 20$ for typical afterglow spectrum. Using
$\langle 1/T \rangle^{-1} = 3.0$days, the total exposure of half a year
is equivalent to a snapshot observation of $85 \times 180 / 3.0 \sim
5000$ square degree. Then, about 2 or 3 orphan afterglows are expected.
Since the variability time scale is only a few days at this magnitude,
continuous monitoring of OGLE III with the time interval of one or two
days is very useful for an afterglow search. The mean redshift
is $10^{\log z} = 0.65$ with a dispersion of $\sigma_{\log z} \sim 0.5$.
This means that a significant fraction of orphan afterglows detectable
by OGLE III should have large redshifts of $z \gtrsim 1$. Then the
brightness of $R \sim 20$ would be much brighter than any kind of 
supernovae and hence orphan afterglows and supernovae can be discriminated.

\subsubsection{ROTSE-III}
The ROTSE group is considering to use their ROTSE-III telescope for an orphan
afterglow search, whose field of view is 3.5 deg$^2$ (Smith et al. 2002;
Kehoe et al. 2002). A search by planned four instruments will cover about
1400 deg$^2$ with a limiting magnitude better than 19th each night
(C. Akerlof 2002, a private communication). This means that ROTSE-III has a
capability of doing an orphan afterglow search equivalent to that
made by SDSS (Vanden Berk et al. 2002) only in one night.  If such a search
is continued for a year, the effective sky coverage could reach $\sim 1400
\times 365 / 1.3 \sim 3.9 \times 10^5$ deg$^2$, taking into account that
$\langle 1/T \rangle^{-1}$ = 1.3 days at this sensitivity.  Then we expect
more than 50 orphan afterglows with $b_{\rm rel} \sim 3$. Since 
$T \lesssim$ 1 day, more than one observations in a night would also be
favored for sufficient time resolution.

\subsubsection{Subaru/Suprime-Cam}

The Subaru Prime Focas Camera (Suprime-Cam) of the 8.2m Subaru telescope is
a unique facility with a field of view (FOV) of $30' \times 30'$, which is more
than 100 times wider the than typical field of views of 8m-class telescopes or
HST.  Therefore this instrument is the most suitable for an orphan afterglow
search at the deepest sensitivities we can achieve.
A sensitivity limit of $R \sim 26$ for point sources\footnote{
This estimate is not seriously affected even if an afterglow is found
in a host galaxy whose luminosity is brighter than the afterglow,
since generally ground-based observation is limited by the sky background,
and in most cases the surface brightness of galaxies is not brighter than
the sky.} is 
achieved by about 10 minutes exposures, and then it is possible to observe
about 10 FOVs in a night with multi-band photometry. 
A clear advantage of the deep sensitivity is that we expect
a large relative beaming factor, $b_{\rm rel} \sim 50$.
Then we expect about 0.4 orphan afterglows. Therefore
a detection is not extremely difficult when systematic
searches are performed using several nights. We show the expected number
assuming a search over a 5-$\rm deg^2$ field in Table \ref{table:projects}.
If detected, a large value of
$b_{\rm rel}$ strongly argues for the existence of many more orphan
afterglows than those associated with GRBs. At this sensitivity,
the time scale of afterglow variability is increased to $10^{\log T}
\sim$ 150 days with $\sigma_{\log T} \sim 0.4$. Therefore, a longer time
scale than supernova searches may be favored.

A problem is discrimination from other transient objects. A few supernovae
are typically found in one FOV of the Suprime-Cam down to $I \sim 25$ (M. Doi
\& N. Yasuda 2002, private communications). At this magnitude, orphan
afterglows may not be sufficiently brighter than supernovae, while the SDSS
transient of Vanden Berk et al. (2002) was about 100 times brighter than the
brightest supernovae.  Most supernovae have thermal spectra that are curved
compared with that of power-law GRB afterglows. Therefore, color-color plots
by multi-band photometry can be used for discrimination (Rhoads 2001), though
it is still uncertain whether this method can remove all supernovae including
peculiar ones. Offset of optical transients from centers of galaxies, close
examination of host galaxies, and past records will be useful to remove AGNs,
as mentioned in the previous section.

\subsubsection{GAIA}

The astrometric satellite GAIA will survey all the sky many times
with a sensitivity of $R \sim 20$. Each location of the sky will be
observed about 40 times separated by more than one month,
during the whole mission (L. Eyer 2002, private
communication). Therefore we expect about 720 orphan afterglows.
While the relative beaming factor is not large ($b_{\rm rel}
\sim 4.3$ at this sensitivity limit), the enormous number expected
might be useful for statistical analysis of orphan afterglow rate.
Because of small $b_{\rm rel}$, the cross check between gamma-ray
observation is crucial to discriminate orphans from ordinary afterglows
associated with observable GRBs.
It is highly desired that a GRB satellite covering a significant
part of all sky is working at the time of GAIA project.

\subsubsection{Dark Matter Telescope}
Even more powerful searches than that by the Subaru/Suprime-Cam would become
possible at the sensitivity of 8m-class telescopes, by the planned Dark
Matter Telescope (DMT)\footnote{http://www.dmtelescope.org}, 
having a 7-deg$^2$ field of view. In the
planned all-sky survey mode, the DMT will cover 20,000 deg$^2$ down to 24th
magnitude twice in the same month, in which about 500 orphan afterglows are
expected with $b_{\rm rel} \sim 20$. On the other hand, the deep probe mode
will cover ten 100-deg$^2$ fields with a sensitivity of 29th magnitude.
Then, in principle, a few thousands of orphan afterglows could be 
detected with $b_{\rm rel} \sim 200$, although it is not yet clear
how they can be discriminated from other transient objects at such a deep 
sensitivity level. It should also be noted that the characteristic 
time scale $T$ is more than 500 days for this sensitivity, and hence
a sufficiently long time interval is required for the search.

\subsection{Radio Observations}

Perna \& Loeb (1998) used faint source counts in radio bands
to constrain the orphan afterglow rate. 
%Here we compare our model with their constraint. 
The radio source counts at 8.44GHz is $\sim 3 \times 10^6$ in all sky for $S
>0.1$mJy (Windhorst et al. 1993; Becker, White, \& Helfand 1995). Following
Perna \& Loeb, we assume that about 3\% of sources at this sensitivity are
variable, yielding an upper bound on the number of orphan afterglows in all
sky as $\lesssim 9 \times 10^4$.  The power index of the radio spectrum is
changing from the initial value of $\delta = -1/3$ (below the peak frequency)
to $\delta \sim 1$ (above the peak frequency), but typically it is $\delta
\sim 0.5$ at a few hundreds days after the burst.  Taking into account this
correction from 8.44 to 5GHz, we obtain $\sim 2 \times 10^3$ orphans expected
by our model. Therefore the current upper bound on
variable radio source counts hardly constrains our model. In fact, the
relative beaming factor expected at this sensitivity is $b_{\rm rel} \sim$ 15,
which is not greater than the maximum expected by the deepest optical search
by Subaru/Suprime-Cam. One might have thought that radio band is potentially
the best waveband for orphan afterglow searches, since radio afterglows are
visible long after the bursts and hence we expect a large $b_{\rm
rel}$. However, in order to achieve $b_{\rm rel} \gtrsim$ 100, the sensitivity
must be better than $\sim$ 1--10 $\mu$Jy.

The difficulty in radio searches is discrimination from other
variable sources such as AGNs. It is difficult to do this
discrimination only in the radio band, but might be possible if we utilize
information in other wavebands.  Intensive examination of variable objects
found in the radio band with cross-checks between the radio, optical, and
X-ray surveys may allow us to find an orphan afterglows in a large ($\sim$
40--50)
sample of variable AGNs at sensitivities of $\sim$ 0.1mJy. 
%The area
%coverage must be wider than $\sim 10 \ \rm deg^2$ to find at least one orphan
%afterglow.

A unique characteristic of radio afterglow, which is different from X-ray or
optical bands, is that $b_{\rm rel}$ does not decrease but stay roughly
constant at $b_{\rm rel} \gtrsim 10$--20 when the search sensitivity is
decreased to $F_\nu \gtrsim$ 1mJy (brightest flux region).  This is coming
from the properties of the radio light curves; the peak of radio light curves
occurs at relatively late time even in the on-axis case. Therefore, the radio
afterglows cannot be detected at large cosmological distances even when an
observer is located on the jet axis, while on-axis early afterglows in X-ray
or optical bands are very bright and they can be detected at almost all the
cosmological distance scale. As can be seen in Fig. \ref{fig:lc-radio}, the
peak radio flux is almost independent of $\thobs$ in some dominant GRBs such
as 991216, 000301c, and 000926. Therefore $b_{\rm rel}$ converges into a
finite value ($\sim$10--20) with $F_{\rm lim} \rightarrow \infty$. 
On the other hand, as mentioned earlier, very
bright on-axis afterglows at large cosmological distances are dominant in the
brightest flux range in the X-ray or optical bands, making $b_{\rm rel}$ to
converge to unity with $F_{\rm lim} \rightarrow \infty$. The sharp
decrease of the mean redshift $\langle z \rangle$ of radio afterglow with
increasing flux in a flux range of $F_\nu \gtrsim$ 1mJy, where $b_{\rm rel}$
is roughly constant, is also consistent with this interpretation (see Fig
\ref{fig:rate-radio}).  This effect then suggests that a relatively shallower,
but wide-field search is an efficient way to constrain the orphan afterglow
rate in radio bands, rather than deeper and narrower searches. In such a
survey, low-redshift afterglows with large $T$ and $\theta_{\rm obs}$ should
be dominant, which may be called as GRB remnants rather than afterglows
(Paczy\'nski 2001; Ayal \& Piran 2001).

Levinson et al. (2002) made a search of orphan afterglows by comparing
the FIRST and NVSS surveys. Their sky coverage is 5990 deg$^2$, and they
found 26 candidates of orphan afterglows, for which they argued that these are
unlikely to be radio supernovae, while the possibility of radio-loud AGNs
cannot be rejected.  The number in 5990 deg$^2$ may be increased to $\sim$ 65
when corrections for incompleteness are made. Their search sensitivity is 6 mJy
at 1.5 GHz, and hence we transformed this into 3.3 mJy at 5 GHz band again
assuming $\delta = 0.5$.  Then we found that our expectation for this search
is $N_{\rm exp} \sim 2.0$ at $b_{\rm rel} \sim 16$. Further observational
inspection of these candidates is necessary, and such effort may reveal some
orphan afterglows.  If, instead, the majority of the candidates turned out to
be orphans, it would indicate a very high orphan afterglow rate, which cannot 
be explained within the jet model with sharp edges.

A future project, Allen Telescope Array (ATA), for the SETI has a
sensitivity of about 0.3mJy at 1.4GHz (0.16mJy at 5GHz if $\delta = 0.5$)
in one minute of integration time, with
a pixel size of $\sim$1 arcmin diameter and a total beam area of 2.5 degree
diameter.  At this integration time, it could cover 25\% of the northern sky
everyday (L. Blitz 2002, a private communication, see also
http://www.seti-inst.edu/science/ata.html).  Although the sky coverage is not
much better than the FIRST/NVSS surveys utilized by Levinson et al. (2002),
the better sensitivity increases the expected number of orphan afterglows 
greatly to $\sim 200$, out to the redshift of
$z \sim 0.6$ with a characteristic variability time scale of 200--300 days. 
Furthermore, ATA will provide radio light
curves sampled everyday, which would be very useful to check whether a
transient source is an orphan afterglow or not.

\section{Discussion}
\label{section:discussion}

\subsection{Comparison with Previous Work}
\label{section:discussion-dalal}

Our result that the detection rate of orphan afterglows of collimated GRBs in
optical bands could be much higher ($b_{\rm rel} \gtrsim 100$ at $R \sim 24$
for GRB 991216) than in the case of spherical GRBs seems apparently in contrast
to that of Dalal et al. (2002), who showed a result that the detection rate
of orphan afterglows is insensitive to the jet opening angle $\thjet$, and
$b_{\rm rel}$ is constant at $\sim$ 3--4 for a search with a sensitivity of
$R \sim 27$. We note that the result of Dalal et al. (2002) was derived
assuming fixed values for the jet-break time and luminosity at the break time
for an on-axis observer. In this case, the maximum viewing angle at which the
afterglow can be detected [$\theta_{\max}$ in eq. (\ref{eq:theta-max})]
%(i.e$.$ the angle for which the peak flux reaches
%the sensitivity limit of the search) 
is proportional to $\thjet$, and hence the
relative beaming factor $b_{\rm rel}$ does not depend on $\thjet$, as can be
seen from equation (\ref{eq:theta-max}).

 However, if the energy of afterglow jets does not vary much among GRBs, as
suggested by Frail et al (2001), despite a wide range of initial jet angles
$\thjet$ (Panaitescu \& Kumar 2001), the flux at the jet-break time $F_0 (\nu,
t_{j,0})$ measured by an on-axis observer is strongly dependent on $\thjet$,
thus, far from the assumption of constancy made by Dalal et al (2002). Within
the framework of relativistic jets, it can be shown that at optical
frequencies $F_0 (\nu, t_{j,0}) \propto \thjet^{-2p}$, where $p$ is the index
of the power-law electron distribution. Furthermore, after the jet-break time
the slope $-\alpha_2$ of the afterglow decay is $-p$ (Rhoads 1999). Then
equation (\ref{fpeak}) leads to a peak flux $F(\nu, t_p)$ for an off-axis
observer that is independent of $\thjet$.  This conclusion can also be reached
by noting that the jet dynamics after the jet-break time is independent of
$\thjet$ (Granot et al. 2002). 
As shown in \S\ref{section:model}, for an off-axis observer the
afterglow light-curve peaks after the jet-break time, therefore the peak flux,
which is determined by the jet dynamics and observer location, does not depend
on the initial jet opening.  That $F(\nu, t_p)$ is independent of $\thjet$
implies that $\theta_{\max}$ does not
depend on $\thjet$ and the original expectation $b_{\rm rel} \propto
(\theta_{\max} / \thjet)^2 \propto \thjet^{-2}$ is restored.

\begin{figure}
\plotone{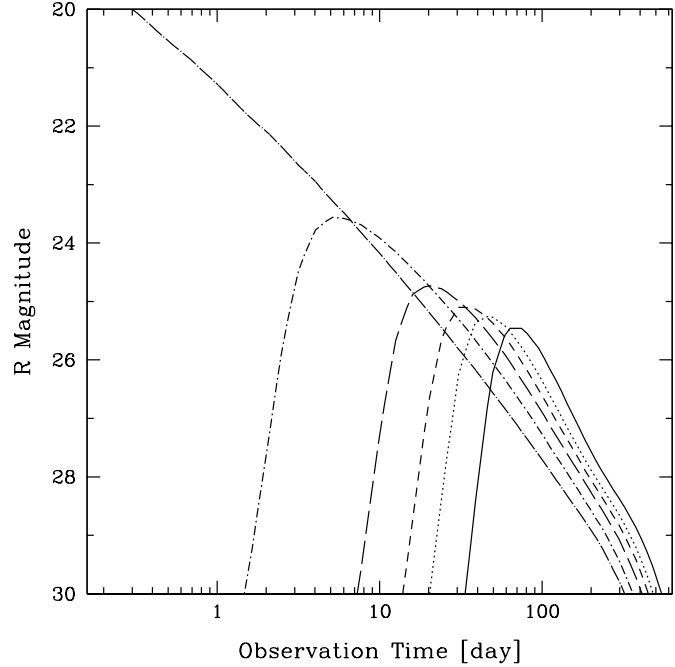}
%\plotone{f8.eps}
\caption{The optical light curves of orphan afterglows,
for a fixed viewing angle $\thobs = 20^\circ$,
but various jet opening angles, $\thjet$.
The solid, dotted, short- and long-dashed, and short- and long-dot-dashed lines
are for $\thjet$ = 1, 2, 3, 5, 10, and 20$^\circ$, respectively.
The redshift is assumed to be 1. Other parameters of the afterglow model
are: $E_{\rm jet} = 3 \times 10^{50}$ erg, $n_{\rm ex} =
1 \rm cm^{-3}$, $\epsilon_e = 0.05$, $\epsilon_B = 3 \times 10^{-3}$, and
$p=2$. }
\label{fig:lc-R-jetangle}
\end{figure}

\begin{figure}
\plotone{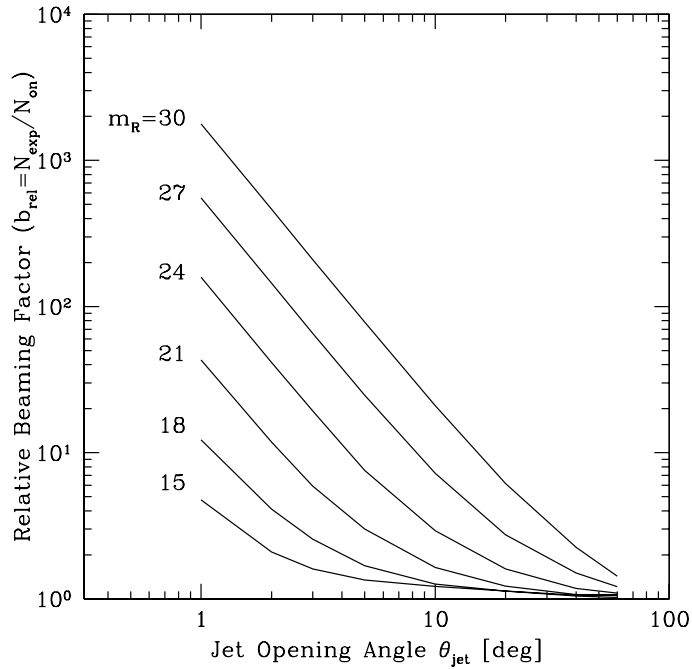}
%\plotone{f9.eps}
\caption{The $\thjet$ dependence of
the relative beaming factor $b_{\rm rel}$, i.e.,
the ratio of rate for all afterglows, including orphans, to 
those associated with prompt GRBs. Different curves correspond
to different search sensitivities in optical band, as indicated in the figure.
Other parameters of the afterglow model
are: $E_{\rm jet} = 3 \times 10^{50}$ erg, $n_{\rm ex} =
1 \rm cm^{-3}$, $\epsilon_e = 0.05$, $\epsilon_B = 3 \times 10^{-3}$, and
$p=2$.}
\label{fig:brel-jetangle}
\end{figure}

The above result regarding the constancy of $F(\nu, t_p)$ is illustrated in in
Figure \ref{fig:lc-R-jetangle}, where we show the light curves of optical
afterglows with various values of $\thjet$ for an observer located at $\thobs
= 20^\circ$. Here we used a fixed set of typical afterglow model parameters
(other than $\thjet$): $E_{\rm jet} = 3 \times 10^{50}$ erg, $n_{\rm ext} = 1
\rm cm^{-3}$, $\epsilon_e = 0.05$, $\epsilon_B = 3 \times 10^{-3}$, and
$p=2$. The peak flux hardly changes with $\thjet$ at $\thjet \lesssim
5^\circ$. Figure \ref{fig:brel-jetangle} shows the beaming factor $b_{\rm
rel}$ as a function of $\thjet$ for various sensitivities. Note that $b_{\rm
rel}$ becomes larger for more collimated GRBs, asymptotically reaching the
expected relation $b_{\rm rel} \propto \thjet^{-2}$ when $\thjet \rightarrow
0$. Thus, orphan afterglow searches could give useful information on the GRB
collimation.

\subsection{Uniform versus Universal Jet Profile}

In this work we have assumed a conical jet with a sharp edge, a uniform
energy per solid angle within the jet opening, and no energy outside (the
{\sl uniform jet model}). Such a model is appropriate if the angular
distribution of the energy has a characteristic angular scale and decreases
rapidly beyond it, e.g., an exponential profile $\propto \exp
(-\theta/\thjet)$. In this model, the observed anticorrelation between
isotropic equivalent luminosity and jet break time can arise if jets have
roughly the same energy but different opening angles $\thjet$ among bursts
(Frail et al. 2001).  However, if the energy per solid angle has a large
variation but no characteristic scale (e.g. a power-law distribution) a
completely different picture is possible. Some recent papers proposed that
the observed luminosity -- break-time anticorrelation can be explained by a
{\sl universal jet} with a non-uniform profile, observed at different viewing
angles (Rossi et al.  2002, Salmonson \& Galama 2002, Zhang \& M\'esz\'aros
2002), which can be an alternative to the uniform jet model. In such a model,
the angular distribution of jet energy per unit solid angle should be a
power-law $(d{\cal E}/d\Omega) \propto \theta^{-2}$ to reproduce the above
mentioned correlation.

It is not straightforward to predict how the orphan afterglow rate is changed
when such a picture is adopted rather than the uniform jet model. If we
have an ideal gamma-ray detector that can detect all GRBs everywhere in the
universe, then we expect that $b_{\rm rel}$ cannot be much greater than 
unity in the universal jet picture, because of the following reason.  We
expect orphan afterglows only when viewing angles larger than an angle
$\theta_\gamma$ corresponding to either the maximum angular spread of the jet
or the angle at which small energy per solid angle and/or the Lorentz factor
yield a barely detectable gamma-ray emission.  The orphan afterglow rate
should then be close to that predicted by our model for a jet of opening
angle $\theta_\gamma$. An estimate of $\theta_\gamma$ can be obtained as
following.  The isotropic equivalent gamma-ray energies calculated by Bloom,
Frail \& Sari (2001) for bursts with known redshifts span 2-3 orders of
magnitude. If the GRB output is mainly determined by the jet energy per solid
angle toward the observer, then the above distribution $(d{\cal E}/d\Omega)
\propto \theta^{-2}$ implies that the dimmest GRB jets are seen at an angle
at least 10 times larger than that for the brightest jets. The latter angle
should be around the smallest jet opening angle of $2^\circ$ found by
Panaitescu \& Kumar (2002) by modeling the broadband emission of ten
afterglows using the uniform jet model\footnotemark,
 \footnotetext{Rossi et al. (2002) have shown that the jet initial opening  
   inferred in the uniform jet model is in fact the observer's angular off-set 
   relative to the jet axis in the universal jet model.}
leading to $\theta_\gamma \gtrsim 20^\circ$. Since the relative beaming
factor decreases with the jet opening angle, as shown in Figure
\ref{fig:brel-jetangle}, we expect that for the universal structured jet
model $\brel \sim$ a few for the reasonable search sensitivities of
$R \lesssim 27$. On the other hand, as we have shown in Fig. 
\ref{fig:rate-R}, it rapidly increases to $b_{\rm rel} \sim 50$
with the sensitivity to $R \sim 26$ in the uniform jet model.

However we may have orphan afterglows due to an insufficient gamma-ray
sensitivity.  Since the gamma-ray luminosity per unit solid angle rapidly
increases with decreasing $\thobs$ in the universal jet model, 
the detection of GRBs at large distances
might be biased toward those with small $\thobs$. If this is the case, we
expect a much higher true GRB rate and large $b_{\rm rel}$ might be possible
also in the universal jet picture.  A test for this case is to see an
anticorrelation between $z$ and $\thobs$, and confirm or reject the $\thobs$
distribution obeying $N(<\thobs) \propto (1 - \cos \thobs)/2$, as predicted
by the universal jet model, for a complete sample of GRBs within a redshift
range. The distribution of $\theta_{\rm obs}$ can be observationally inferred
from either luminosity function of GRBs with $z$ measurements or jet break
time obtained by afterglow light-curve fitting.  However, the present sample
of GRBs with known redshifts is too small and we must await future
observations.

Another test possible by orphan afterglow observation is to examine the early
behavior of orphan afterglow light curves.  For $\thobs > \thjet$, in the
uniform jet model we always expect no afterglow flux at the earliest
stage, and the afterglow flux should show gradual increase until the peak
flux time, $t_p$, given in \S \ref{section:model}. When $b_{\rm rel} \gg 1$,
the orphan afterglow events should be dominated by such cases, and hence we
expect that the majority of orphan afterglows should show a slow rise at the
beginning.  On the other hand, the X-ray and optical emission should start
immediately after the prompt burst when $\thobs < \theta_\gamma$ in the
universal jet picture, since there is ejecta moving toward the observer.
The initial slow rise of light curves is possible only when
$\thobs > \theta_\gamma$, but such events should be relatively rare because
of the large $\theta_\gamma$ inferred from observations and low absolute
luminosity for the large $\thobs$ cases. The detection of such early rise of
light curves may not be easy even in the case of the uniform jet model
because the rising time scale is still smaller than the overall variability
time scale $T$ as can be seen in Figs. \ref{fig:lc-X} and \ref{fig:lc-R}.
If detected, however, it would argue for the uniform jet picture.

\subsection{Caveats of Our Predictions}
\label{section:discussion-caveats}

Here we describe several caveats of our prediction which should be
kept in mind when one compares it to observed data.

It is theoretically conceivable that the GRB central engine ejects not only
the ultrarelativistic outflow that is responsible for GRBs, but also less
relativistic matter with comparable total energy.  Prompt gamma-ray emission
may be very dim or completely absent from such less relativistic components
of ejecta, while the afterglow emission similar to those associated with GRBs
might be possible.  Even if the ultrarelativistic component may be strongly
collimated to produce beamed GRBs, more isotropic, less relativistic
component could be associated to most of GRBs. It is also possible that there
are much greater number of events of dirty fireball or failed GRBs without
prompt gamma-ray emission to any direction, than that of observed GRBs. 
The brightest class of core-collapse
supernovae, called hypernovae (Iwamoto et al. 2000; Nakamura et al. 2001;
Mazzali et al. 2002), might also be failed GRBs.
In either case, the orphan afterglow rate can be
increased significantly from our prediction.  Even if an orphan afterglow is
discovered, it is not an easy task to discriminate such another component of
ejecta or failed GRBs from the pure effect of GRB jet collimation (Huang,
Dai, \& Lu 2002).

 One possible way of discrimination between an orphan afterglow due solely to
the viewing geometry from one due to a "dirty" fireball is provided by the
afterglow decay rate. In the former case, the jet collimation should yield a
decay slope $\alpha \sim -2$, while a less collimated component or a failed
GRB are expected to yield $\alpha \sim -1$. The slower rise of the
light-curves of off-axis, collimated jets is also useful as discussed in the
previous subsection (see also Huang et al. 2002).  Stronger linear
polarization is also expected for off-axis, collimated jets (Sari 1999;
Ghisellini \& Lazzati 1999; Granot et al. 2002).  Finally, the wavelength and
sensitivity dependence of $b_{\rm rel}$ might be able to discriminate the
above possibilities. The signature of off-axis afterglows would be the
continuous increase of $b_{\rm rel}$ with the search sensitivity, though this
trend might be mimicked by failed GRBs if their event rate is continuously
increasing with decreasing energy output to the dirty component. The
constraint of $\brel \lesssim$ several by the past X-ray searches already
indicates that the event rate of failed GRBs producing X-rays cannot be much
higher than that of successful GRBs. On the other hand, the so-called X-ray
rich GRBs might be a population between the failed and ordinary GRBs (e.g.,
Kippen et al. 2002).

 We have assumed that the sky coverage of the gamma-ray emission is that 
corresponding to the jet opening angle obtained from afterglow modeling.
However, this is not necessarily warranted. The causally connected angular
scale during the prompt GRB phase is only $\theta \sim \Gamma^{-1}$, which
should be much smaller than the jet opening angle $\thjet$ (Kumar \& Piran
2000). Then it is possible that the jet outflow, and thus its gamma-ray
emission, is inhomogeneous on an angular scale much smaller than $\thjet$,
while the afterglow emission arising later in the jet evolution is more
homogeneous. In this case the GRB rate is significantly underestimated, as
there could be a number of undetected GRBs due to some low-energy patches
moving toward the observer, even if the observer is located within the jet
opening angle.

 We have also assumed that the ten GRB afterglows used here are
representative for all varieties of GRB afterglows. Although X-ray afterglows
were observed for the majority of GRBs, optical afterglows were found for
less than half of GRBs.  Some afterglows were missed because of late
follow-up observations, but there were also unusually dim optical afterglows
(Fynbo et al. 2001; Lazzati, Covino \& Ghisellini 2002). Several explanations
can be considered including dust extinction and very large redshift. Recent
studies of high redshift galaxies suggest that about half of stars are formed
in very dusty galaxies, the fraction being much larger at high redshift than
in the local universe (see, e.g., Totani \& Takeuchi 2002; Ramirez-Ruiz,
Trentham, \& Blain 2002). The sample of the ten afterglows used here is
clearly biased toward afterglows occurring in less dusty galaxies. Then our
estimate of optical orphan afterglow rate may be overestimated by a factor of
about two. When a GRB occurs in a molecular cloud or high density region with
significant amount of dust, the dust along the direction of the jet might be
destroyed by strong optical-UV flash (Waxman \& Draine 2000) or early X-ray
radiation (Fruchter, Krolik, \& Rhoads 2001), making optical afterglows
visible for on-axis observers. However, this effect should be small for
off-axis observers, because the off-axis X-ray afterglow flux is much weaker
when the light-curve peaks.  This phenomenon may effectively reduce the
observed relative beaming factor $b_{\rm rel}$.  Finally, it should be noted
that the prediction for radio orphan afterglows is not affected by dust
extinction.

\section{Conclusions}
\label{section:conclusions}

We presented a quantitative prediction for the detection rate of orphan GRB
afterglows, based on one of the latest afterglow models that has been tested
with a number of observed afterglows.  We found that the orphan afterglow rate
sensitively depends on afterglow model parameters, and a fairly large $b_{\rm
rel}$ ($\gtrsim $100) is possible for some types of GRBs, by an optical search
with reasonable depth ($R \gtrsim 24$).  We derived our best-guess prediction
of orphan afterglow rate, by taking a weighted mean of the ten sets of
afterglow parameters that fit to ten well-observed afterglows.  Although
there are a number of effects or caveats which could significantly change our
predictions, the prediction will be useful as ``a baseline model'' when we
interpret the results of past and future surveys for extragalactic transient
objects.

Our prediction is consistent with all the past surveys in X-rays, optical,
and radio wavebands. Greiner et al. (2000) reported that there are about 10
possible candidates of orphan afterglows, which is interestingly very close
to our expectation. Although the SDSS transient reported by Vanden Berk et
al. (2002) is very likely to be a radio-loud AGN, their search has already
reached a meaningful sensitivity, since our expectation is about 0.2
afterglows for this search.  A recent search by Levinson et al. (2002) found
about 30 candidate radio afterglows by comparing the FIRST and NVSS surveys,
while our model expects about two orphans.

Detection of orphan afterglows seems not extremely difficult in future
surveys. Accumulation of data of advanced X-ray satellites such as Chandra
and XMM-Newton might detect orphan afterglows whose rate is enhanced by a
relative beaming factor of $b_{\rm rel} \sim 40$ compared with that for
GRB-associated afterglows. OGLE III and ROTSE-III projects could detect
a few and a few tens of orphan afterglows with $b_{\rm rel} \sim $ 3--4 in
half a year at $R \sim 20$ and 19, respectively, 
providing nightly light curves.
The southern SDSS observation ($R \sim
23$) could detect about 40 orphan afterglows with $b_{\rm rel} \sim 14$
during five-year operation.
Further deep optical surveys by
Subaru/Suprime-Cam ($R \sim 26$) might detect an orphan afterglow with $b_{\rm
rel} \sim 50$. In the more distant future,
GAIA could detect $\sim$ 900 afterglows 
down to $R \sim 20$, and DMT could detect 500 and 2000 orphans at
$R = 24$ and 29, respectively.
%Intensive examination of variable radio sources down to
%$\sim$ 0.1mJy might reveal an orphan afterglow
%among about 40-50 variable AGNs. 
ATA would detect about 200 orphans
with $b_{\rm rel} \sim 15$ at $\sim$0.3mJy (1.4GHz) in the radio band.
Further effort for these searches is encouraged
in the near future. Finally, we should make a point that a future GRB mission
monitoring a significant part of all sky, like BATSE, is desired to check
whether a candidate orphan afterglow is really an ``orphan''. It is crucial
especially for shallow searches of orphan afterglows with small relative
beaming factor.

\acknowledgments{ We would like to thank C. Akerlof, L. Blitz, M. Doi,
L. Eyer, \v{Z}. Ivezi\'c, D. Lazzati, T. Mihara, T. Murakami, B. Paczy\'nski,
E. Ramirez-Ruiz, D. Vanden Berk, and N. Yasuda for many useful comments and
discussions.  T.T. has been financially supported by the JSPS Postdoctoral
Fellowship for Research Abroad. A.P. acknowledges the support
received from the Lyman Stizer, Jr. Fellowship.
}

%%%%%%%%%%%% References %%%%%%%%%%%%%%%%%%

\begin{table}
%\tiny
\caption{Expected Number of Afterglows in Various Surveys}
\begin{tabular}{ccccccccccc}
\hline \hline 

Survey Name & Sensitivity & Area [deg$^2$] &
$N_{\rm exp}$ & $N_{\rm on}$ & $
N_{\rm exp}/N_{\rm on}$ & $10^{\langle \log z
\rangle}$  & $\sigma_{\log z}$ & $10^{\langle
\log T \rangle}$  & $\sigma_{\log T}$ &
$\langle 1/T \rangle^{-1}$ 

\\ 

\hline

\multicolumn{11}{c}{X-ray Observation} \\ 

\hline 

ROSAT\tablenotemark{a} & $1 \times 10^{-12}$ & $9.4 \times 10^4$ & 8.3 & 2.8
& 3.0 & 0.93 & 0.43 & 1.5 & 0.52 & 0.81 \\ 

\hline 

\multicolumn{11}{c}{Optical Observations} \\ 

\hline 

SDSS 1\tablenotemark{b}  & 19 & 1500 & 0.20 & 0.067 & 3.0 & 0.62 & 0.50 & 3.5 & 0.68 & 1.3 \\

SDSS 2\tablenotemark{c}  & 19 & 10000 & 1.3 & 0.45  & 3.0 & 0.62 & 0.50 & 3.5 & 0.68 & 1.3 \\

ROTSE-III  & 19 & $3.9 \times 10^5$ & 53 & 18  & 3.0 & 0.62 & 0.50 & 3.5 & 0.68 & 1.3 \\

OGLE III  & 20 & 5000          & 2.2 & 0.51 & 4.3 & 0.65 &
  0.48 & 6.8 & 0.62 & 3.0 \\

GAIA  & 20 & $1.7 \times 10^6$ & 720 & 170 & 4.3 & 0.65 &
  0.48 & 6.8 & 0.62 & 3.0 \\

Schaefer\tablenotemark{d}  & 21 & 1500 & 1.8 & 0.29 & 6.2 & 0.69 & 0.46 & 12 & 0.58 & 5.7 \\

SDSS 3\tablenotemark{e}    & 23 & 4500 & 36 & 2.6 & 14 & 0.77 & 0.42 & 32 & 0.51 & 18 \\

Supernova\tablenotemark{f}  & 23 & 200 & 1.6 & 0.11 & 14 & 0.77 & 0.42 & 32 & 0.51 & 18 \\ 

DMT 1\tablenotemark{g}  & 24 & 20000 & 480 & 21 & 23 & 0.81 & 0.39 & 58 & 0.47 & 36 \\ 

Subaru & 26 & 5    & 0.73 & 0.013 & 55 & 0.93 & 0.32 & 150 & 0.40 & 97 \\ 

DMT 2\tablenotemark{h}  & 29 & 1000 & 2100 & 11 & 190 & 1.3 & 0.23 & 670 & 0.32 & 460 \\ 

\hline

\multicolumn{11}{c}{Radio Observation} \\ 

\hline 

FIRST/NVSS\tablenotemark{i}  & 3.3 & 5990 & 2.0 & 0.13 & 16 & 0.13 & 0.32 & 220 & 0.39 & 140 \\ 

ATA  & 0.16 & 5200 & 210 & 14 & 15 & 0.63 & 0.32 & 250 & 0.38 & 160 \\ 

\hline \hline
\end{tabular}
\\

\tablecomments{ 
Col. (2): Sensitivity in $\nu F_\nu / ({\rm erg \ cm^{-2} s^{-1}})$ at 1 keV
for the X-ray band, in $R$ magnitude for the optical band, and $F_\nu$/mJy at
5GHz in the radio band. The sensitivity is corrected based on a typical
afterglow spectrum when observation was made in a slightly different band.
Col. (3): Area of a snapshot observation. When a survey is a consecutive
monitoring longer than the typical afterglow time scale $T$, we converted the
exposure (area$\times$time) into an equivalent area of a snapshot
observation using $\langle 1/T \rangle$. (See text for detail.)
Col. (4):The total number of all GRB afterglows, including orphans,
expected to be detectable by a snapshot observation with the surface are
shown in the third column.
Col. (5): The same as Col. (4), but only for 
afterglows associated with prompt gamma-ray emission.
Col. (6): The relative beaming
factor $b_{\rm rel} \equiv N_{\rm exp}/N_{\rm on}$.  Cols. (7, 8): The mean and
1$\sigma$ dispersion of $\log z$.  Cols. (9, 10): 
The mean and 1$\sigma$ dispersion of $\log T$, where
$T$ is the time duration over which afterglows are brighter than the
sensitivities, in units of days. Col. (11): 
The mean of $T^{-1}$.}

\tablenotetext{a}{The ROSAT All Sky Survey (Greiner et al. 2000).}
\tablenotetext{b}{A search by
Vanden Berk et al. (2002) using the first 1500deg$^2$ field of the SDSS data.}
\tablenotetext{c}{
The same as SDSS 1, but after the completion of the SDSS project.}
\tablenotetext{d}{
  A search made by Schaefer (2002).}
\tablenotetext{e}{
  The number expected in the southern
  sky after the completion of the SDSS project.}
\tablenotetext{f}{
  The number expected
   in past supernova surveys.}
\tablenotetext{g}{The all-sky survey mode of the Dark Matter Telescope.}
\tablenotetext{h}{The deep probe mode of the Dark Matter Telescope.}
\tablenotetext{i}{
 The search made by Levinson et al. (2002)
 by using the FIRST and NVSS surveys.}

\label{table:projects}
\end{table}

%%%%%%%%%%%%%%% Figures %%%%%%%%%%%%%%%%%%%%%%

\epsscale{0.8}

\epsscale{0.8}

\epsscale{0.8}

\epsscale{0.5}

\end{document}